\documentclass[conference]{IEEEtran}
\IEEEoverridecommandlockouts
\usepackage{cite}
\usepackage{amsmath,amssymb,amsfonts}
\usepackage{algorithmic}
\usepackage{graphicx}
\usepackage{textcomp}
\usepackage{xcolor}
\usepackage{booktabs} 
\usepackage{tabularx} 
\usepackage{comment}
\usepackage{subcaption}

\begin{document}

\title{Information Transmission in Quorum Sensing for Gut
Microbiome
}

\author{O. Tansel Baydas,~\IEEEmembership{Student Member,~IEEE,} Efe Yatgin,
        Ozgur~B.~Akan,~\IEEEmembership{Fellow,~IEEE}
\thanks{The authors are with  the Internet of Everything (IoE) Group, Electrical Engineering Division, Department of Engineering, University of Cambridge, Cambridge CB3 0FA, UK 
(e-mails: efe.yatgin@gmail.com, \{otb26, oba21\}@cam.ac.uk).}
\thanks{O. B. Akan is also with the Center for neXt-generation Communications (CXC), Department of Electrical and Electronics Engineering, Ko\c{c} University, Istanbul 34450, Turkey (e-mail: akan@ku.edu.tr).}}

\maketitle

\begin{abstract}
Microorganisms employ sophisticated mechanisms for intercellular communication and environmental sensing, with quorum sensing serving as a fundamental regulatory process. Dysregulation of quorum sensing has been implicated in 
various diseases. While most theoretical studies focus on mathematical modeling of quorum sensing dynamics, the communication-theoretic aspects remain less explored. In this study, we investigate the information processing capabilities of quorum sensing systems using a stochastic differential equation framework that links intracellular gene regulation to extracellular autoinducer dynamics. We quantify mutual information as a measure of signaling efficiency and information fidelity in two major bacterial phyla of the gut microbiota: Firmicutes and Bacteroidetes.
\end{abstract}


\section{Introduction}

The human gut microbiome is dominated by \textit{Firmicutes} and \textit{Bacteroidetes}, representing over 90\% of microbial cells \cite{kostic2014microbiome}. Their numerical and metabolic influence has made them central in microbiome studies, with \textit{Eubacterium rectale} (Firmicutes) and \textit{Bacteroides thetaiotaomicron} (Bacteroidetes) as representative species that influence metabolism and immune homeostasis. Gut species coordinate behavior via quorum sensing (QS) molecules \cite{zhang2024quorum}, particularly the universal autoinducer-2 (AI-2) \cite{bassler1994multiple}, synthesized by LuxS \cite{vendeville2005making}. AI-2 activity is consistently detected in the intestine \cite{sperandio2003bacteria}. However, QS contributions differ: LuxS and AI-2 synthase occur in 83.33\% of Firmicutes but only 16.3\% of Bacteroidetes \cite{thompson2015manipulation}. Alterations in the Firmicutes/Bacteroidetes (F/B) ratio have been proposed as a biomarker for human health. The F/B ratio has been studied in inflammatory bowel diseare (IBD) \cite{yin2025progress} and hypertension \cite{yang2015gut}.

The study of microbial communication and information processing has both theoretical and applied contributions. Transmission efficiency and capacity in cellular signaling define limits imposed by biochemical noise and memory effects \cite{sarkar2023efficacy, sarkar2020sparse}. Alternative signaling modes, such as electromagnetic mechanisms, may provide higher capacity than traditional QS \cite{barani2025electromagnetic}. Bacteria also integrate mechanical sensing with chemical signals to regulate behavior \cite{dufrene2020mechanomicrobiology}. \cite{ tkavcik2016information, mukherjee2019bacterial} have shown how microbial populations integrate information to coordinate collective behavior under diverse conditions.

\begin{figure}[htp!]
    \centering
    \includegraphics[width=1\linewidth]{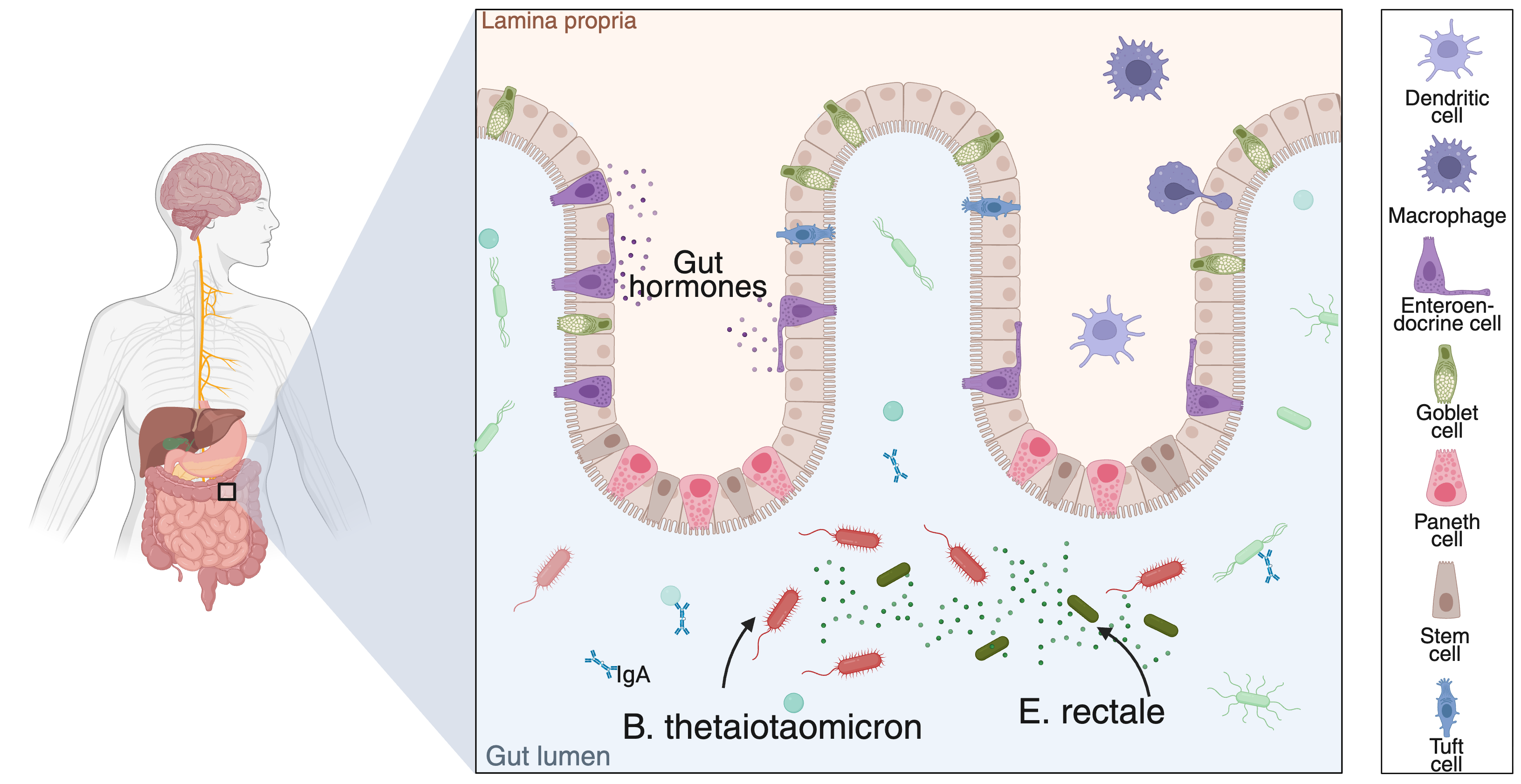}
    \caption{Human gut and its major cells and microorganisms}
    \vspace{-1.76em}
    \label{fig:gut_figure}
\end{figure}

QS-based coordination has been analyzed to identify ideal sensing strategies \cite{taillefumier2015optimal}. These insights connect to ecological and metabolic models simulating microbial interactions, competition, and multistability in the gut microbiome \cite{adrian2023mathematical, shoaie2013understanding}.

Most existing studies on QS in the gut microbiome either employ deterministic modeling or focus on correlating microbial abundances with health outcomes, which overlook two critical points: (i) the inherent stochasticity of biochemical signaling and (ii) the quantitative limits of information transmission in noisy environments. While stochastic models have been applied in other microbial contexts, their integration with information-theoretic metrics remains rare in gut microbiota studies. Furthermore, existing analyses often treat QS as an on/off switch rather than quantifying reliability. 
This lack of a quantitative understanding limits connecting QS dynamics to shifts in microbial balance observed in diseases. 

Beyond systems biology, microbial communication is framed within the Internet of Bio/Nano Things (IoBNT) and Internet of Microbial Colonies (IoMC). IoBNT envisions biological entities or nanodevices as nodes in bio-cyber networks, where molecular communication (MC) underlies information exchange \cite{akyildiz2015internet}. IoMC extends this by modeling microbes as interconnected information-processing networks, emphasizing QS and metabolite exchange \cite{koca2024bacterial}. Embedding gut microbial interactions in these frameworks underscores their relevance for diagnostics and bio-inspired communication systems.

\textbf{To address these gaps}, we develop a stochastic differential equation (SDE) framework that incorporates intrinsic and extrinsic noise into QS dynamics between two gut bacteria, Eubacterium rectale (Firmicutes) and Bacteroides thetaiotaomicron (Bacteroidetes). Unlike prior works, our model couples intracellular monitor protein dynamics with autoinducer production, providing a more realistic characterization of microbial communication. 
We evaluate how regulatory architectures and and noise sources influence the capacity of QS channels. 

\section{System Model}

QS in the gut can be viewed as an MC channel, where producing cells encode population state into autoinducer concentrations and sensing cells decode signals to coordinate behavior. The illustration of human gut is given Fig.\ref{fig:gut_figure}. In a noisy biochemical environment, the key question is not signal magnitude but how reliably levels can be distinguished. Mutual information provides a rigorous measure of signaling fidelity, computed here using an SDE framework linking intracellular regulatory protein dynamics with extracellular autoinducer production through deterministic feedback and stochastic fluctuations.

\paragraph{Density Definitions}

We denote by \(\rho_B(t)\) and \(\rho_F(t)\) the population densities of \emph{Bacteroides thetaiotaomicron} (\(B\)) and \emph{Eubacterium rectale} (\(F\)), respectively. We treat \(\rho_B(t)\) and \(\rho_F(t)\) as externally prescribed time-dependent functions. That is, we assume the microbial population dynamics are either known or controlled experimentally. This assumption allows us to focus on how population size modulates and responds to quorum-sensing behavior. A commonly used family of functions for modeling density growth is a multiplicative exponential ramp:\vspace{-0.5em}
\[
\rho_B(t) = \rho_{-,B} \left( \frac{\rho_{+,B}}{\rho_{-,B}} \right)^{t/T_B},
\quad
\rho_F(t) = \rho_{-,F} \left( \frac{\rho_{+,F}}{\rho_{-,F}} \right)^{t/T_F},
\]

where $0 \le t \le T_B, T_F.$ Here,
\(\rho_{-,B}, \rho_{-,F}\) are the initial densities of species \(B\) and \(F\), \(\rho_{+,B}, \rho_{+,F}\) are the final densities, \(T_B, T_F\) represent the duration of population expansion. This form assumes exponential-like (log-linear) growth in log-density space, reaching to \(\rho_+\). By defining these externally, one could model growth using ramps (as above), step functions, logistic curves, or even piecewise experimental data.

\paragraph{Monitor Protein (MP) Dynamics} 

\begin{figure}[htp!]
    \centering
    \includegraphics[width=0.65\linewidth]{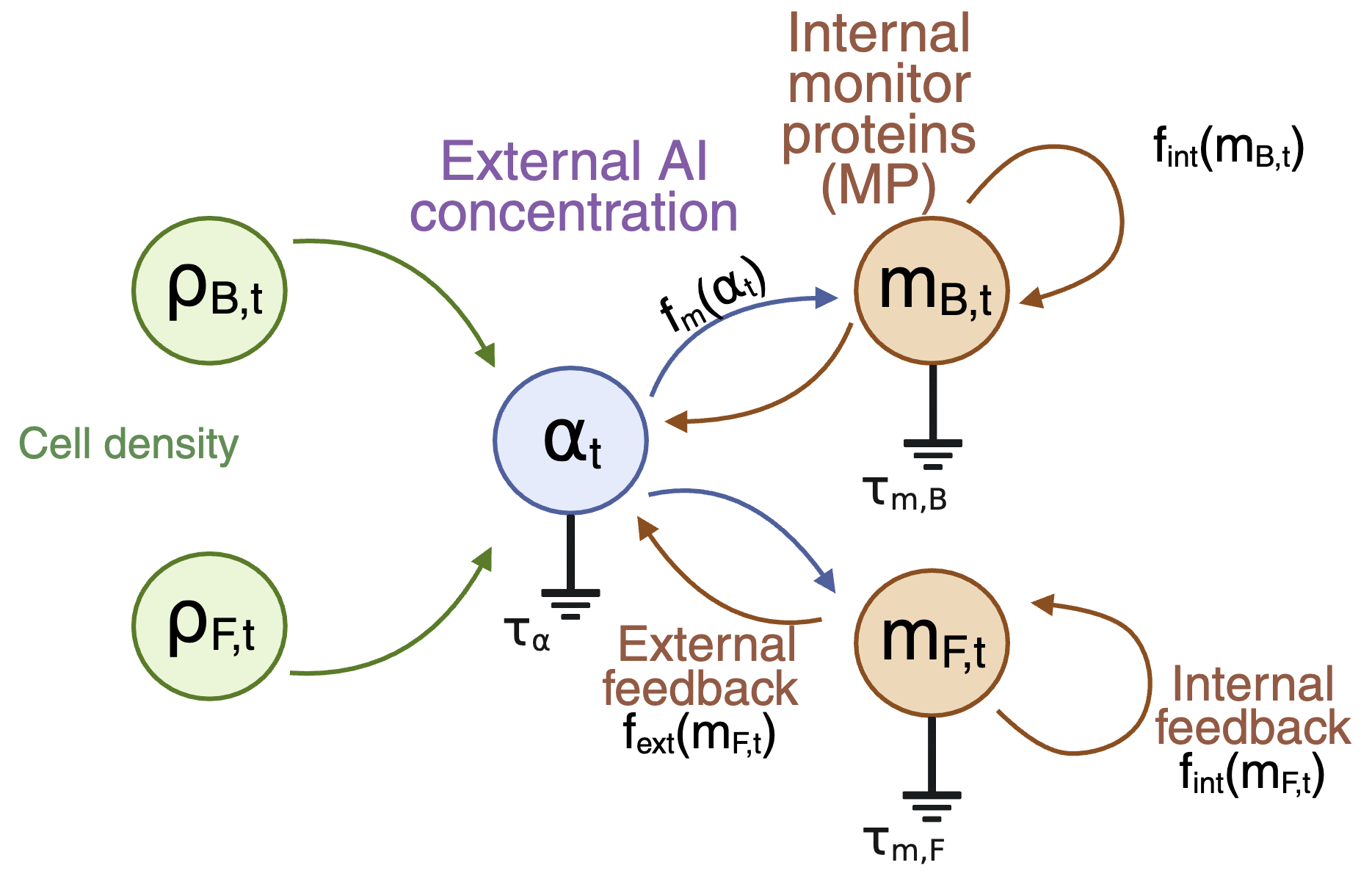}
    \caption{Dependency modeling quorum sensing with internal and external feedback from MP expression to AI signaling.}
    \vspace{-1.5em}
    \label{fig:qs_circuit}
\end{figure}

Each bacterium maintains an intracellular MP that integrates information from the shared extracellular autoinducer \(a(t)\). The dependancy diagram of QS is given in Fig.\ref{fig:qs_circuit}. MP acts as a biochemical proxy for population-wide signaling activity and governs each cell’s response to QS. The internal MP abundance in cell \(i\) of species \(B\) is denoted \(m_{B,i}(t)\), and in cell \(j\) of species \(F\) as \(m_{F,j}(t)\). These MP levels evolve according to coupled SDEs with drift and noise components \cite{taillefumier2015optimal}:
\vspace{-2mm}
\begin{equation}
\begin{aligned}
d\,m_{s,k}(t)
&= -\tfrac{m_{s,k}(t)}{\tau_{m}^{(s)}(\rho_B,\rho_F)}\,dt \\
&\quad + \mu^{(s)}(\rho_B,\rho_F)\,
        f_m^{(s)}\!\bigl(a(t), m_{s,k}(t)\bigr)\,dt \\
&\quad + \sqrt{2\sigma_{m}^{(s)}}\,dW_t^{(k,s)}.
\end{aligned}
\end{equation}

Here, \(s \in \{B,F\}\) denotes the species, and \(k\) indexes individual cells. The first term in equation represents effective MP degradation, which includes both true molecular degradation and dilution from cellular growth. The degradation time constant depends on the total density:
$\tau_m^{(i)}(\rho_B,\rho_F) = 1/(\delta^{(i)} + \mu_g(\rho_B+\rho_F)),$
where $\delta^{(i)}$ is basal degradation rate of the
MP in species \(i\), \(\mu_g(\rho)\) is density-dependent growth rate that dilutes intracellular species. If growth dilution is negligible, a simpler form \(\tau_m^{(i)} = 1/\delta^{(i)}\) may be used. The second term in the drift is a regulated production term, scaled

by \(\mu^{(i)}(\rho_B, \rho_F)\). Production depends nonlinearly on the sensed AI concentration $a(t)$ and is modeled by a Hill function:
$f_m^{(i)}(a,m) = k_{\mathrm{resp}}^{(i)}\, a^n/(K_A^n + a^n),$
where \(k_{\mathrm{resp}}^{(i)}\) is maximum MP synthesis rate in species \(i\), \(K_A\) is half-saturation constant, \(n\) is Hill coefficient controlling cooperativity. This form captures saturable activation of MP production as AI levels rise. The stochastic term \(\sqrt{2\sigma_{m}^{(i)}}\,dW_t^{(i)}\) introduces intrinsic cell-to-cell variability. This reflects biological sources of noise such as transcriptional bursting, fluctuations in translation, and molecular degradation. 

The MP can represent a synthetic reporter, transcriptional regulator, or any QS-regulated gene product. It actively participates in feedback and signaling, with dynamics coupled through the shared AI pool: increases in \(m_{B,i}\) raise AI, which in turn elevates \(m_{F,j}\), and so on.

\paragraph{Autoinducer (AI) Dynamics} 

Both bacterial species contribute to and respond to a shared extracellular signaling molecule, AI. Let \(a(t)\) denote the concentration of this AI at time \(t\). This variable mediates intercellular communication, coordinates collective behavior, and enables cross-species regulation. The dynamics of \(a(t)\) are governed by a SDE that includes deterministic production, degradation terms, and extrinsic noise:
\begin{equation}
\begin{aligned}
d\,a(t)
&= -\tfrac{a(t)}{\tau_a}\,dt
   + \rho_B(t)\,\langle f_{\mathrm{ext}}^{(B)}(m_{B,i}(t)) \rangle_i\,dt \\
&\quad + \rho_F(t)\,\langle f_{\mathrm{ext}}^{(F)}(m_{F,j}(t)) \rangle_j\,dt \\
&\quad + \sqrt{2\,\tfrac{\rho_B(t)+\rho_F(t)}{V}\,\sigma_a^2}\,dW_t
\end{aligned}
\end{equation}

The term $\displaystyle -\tfrac{a(t)}{\tau_a}$ models first-order autoinducer (AI) decay by dilution, enzymatic degradation, or gut washout. The decay time is $\tau_a = 1/(k_{\mathrm{out}} + \mu_{\mathrm{gut}})$, with $k_{\mathrm{out}}$ the environmental loss rate (diffusion or degradation) and $\mu_{\mathrm{gut}}$ the clearance from mucus turnover or peristalsis. The second and third terms give AI production by each species as population density times mean per-cell secretion 
$\rho_s(t)\,\langle f_{\mathrm{ext}}^{(s)}(m_{s,i}(t)) \rangle_i$ \cite{taillefumier2015optimal}. 
The functions $f_{\mathrm{ext}}^{(B)}(m)$ and $f_{\mathrm{ext}}^{(F)}(m)$ describe AI dependence on MP levels; in the simplest case, they are constant: 
$f_{\mathrm{ext}}^{(B)}(m)\equiv \alpha_{\mathrm{LuxS}}^{(B)}, \ f_{\mathrm{ext}}^{(F)}(m)\equiv \alpha_{\mathrm{LuxS}}^{(F)}$, 
with fixed rates $\alpha_{\mathrm{LuxS}}^{(B)}$ and $\alpha_{\mathrm{LuxS}}^{(F)}$. The total secretion is density times this rate, yielding deterministic drift $\rho_B(t)\,\alpha_{\mathrm{LuxS}}^{(B)} + \rho_F(t)\,\alpha_{\mathrm{LuxS}}^{(F)}$, so AI accumulates additively in the medium, independent of origin.

The final term accounts for stochastic fluctuations. This extrinsic noise reflects heterogeneity in secretion and environmental randomness, modeled by a standard Wiener process $W_t$. Fluctuations scale with population size through a square-root dependence, consistent with central-limit aggregation.

AI concentration $a(t)$ serves as the shared signal: each cell decodes it into $m(t)$, which regulates MP levels and feeds back into AI secretion. This creates a nonlinear loop where AI dynamics integrate contributions from both populations.

In simulations, densities $\rho_B(t)$ and $\rho_F(t)$ were updated first, then $a(t)$, followed by all $m_{B,i}(t)$ and $m_{F,j}(t)$. This sequence was iterated until equilibrium or the desired transient was reached. Model parameters, simulation settings, and initial conditions are listed in Table \ref{tab:model_params}.

\begin{table}[htp!]
\vspace{-1em}
\centering
\begin{tabularx}{1.01\linewidth}{|l|c|l|c|}
\hline
\textbf{Model Param.} & \textbf{Value} & \textbf{Sim. / Init.} & \textbf{Value} \\
\hline
$\rho_-$      & $1.0 \times 10^{-4}$              & $N_{\text{cells}}$ & 100 \\
$\rho_+$      & $1.0$                             & $dt$               & 0.01 \\
$T_{\text{div}}$ & 1800.0                         & $t_{\text{max}}$   & 1800.0 \\
$\gamma$      & 0.0                               & $\Omega_0$         & 0.1 \\
$v_{\text{avg}}$ & $1.0 \times 10^{-18}$          & $V_{\text{sys}}$   & $1.0 \times 10^{-12}$ \\
$\tau_{\delta}$  & 1800.0                         & $\tau_{a}$         & 10 \\
\hline
\multicolumn{2}{|l|}{\textbf{Firmicutes- Bacteroidetes (F/B) ratio )}} & \multicolumn{2}{c|}{7 / 3} \\
\hline
\end{tabularx}
\caption{Summary of model parameters, simulation settings, and initial conditions.}
\vspace{-1em}
\label{tab:model_params}
\end{table}

\section{Results and discussion}

\begin{figure}[htp!]
    \vspace{-0.5em}
    \centering
    \begin{subfigure}[b]{0.49\linewidth}
        \vspace{-0.5em}
        \centering
        \includegraphics[width=\linewidth]{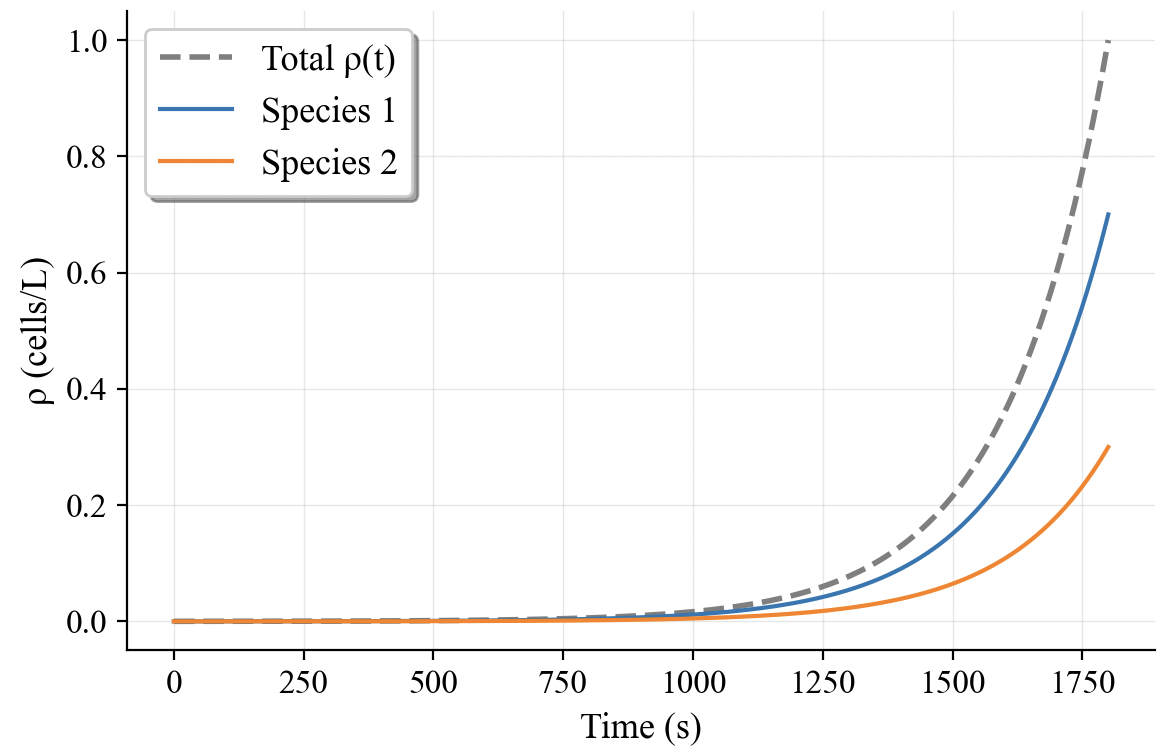}
        \caption{Population dynamics of Firmicutes and Bacteroidetes.}
        \label{fig:population}
    \end{subfigure}
    \hfill
    \begin{subfigure}[b]{0.49\linewidth}
        \vspace{-0.5em}
        \centering
        \includegraphics[width=\linewidth]{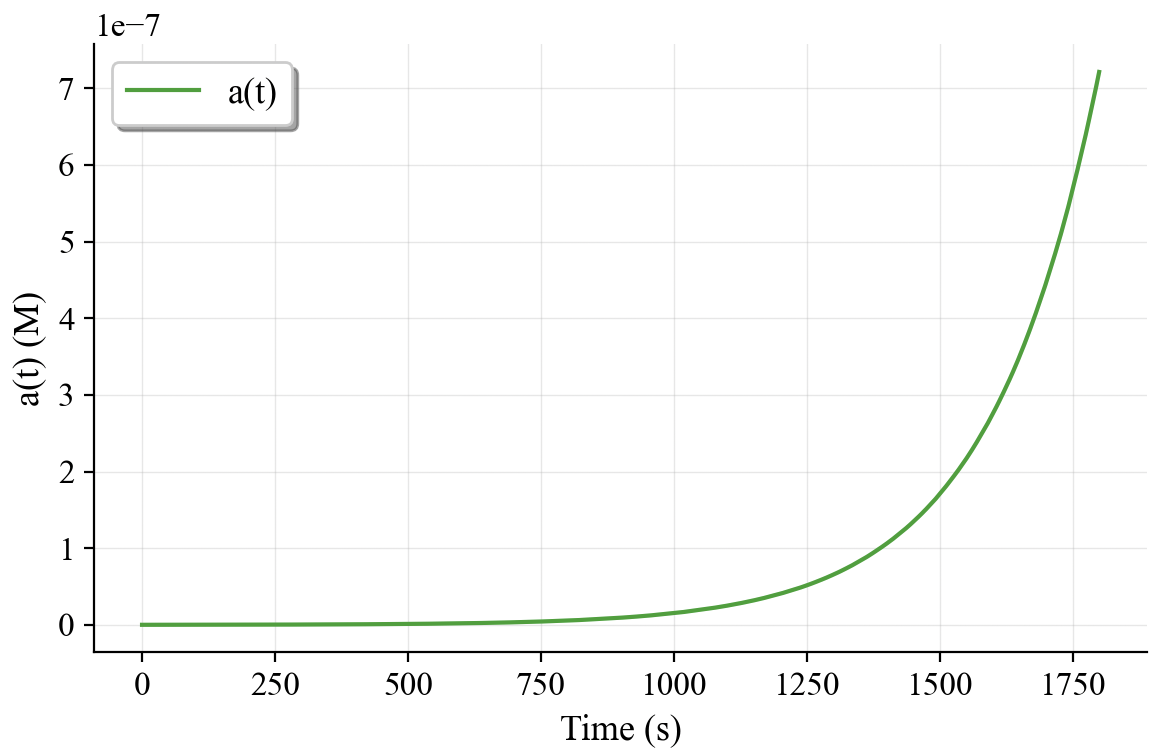}
        \caption{Extracellular shared autoinducer concentration.}
        \label{fig:autoinducer}
    \end{subfigure}
    \caption{Population growth and autoinducer dynamics.}
    \vspace{-1em}
    \label{fig:pop_auto}
\end{figure}

Fig.\ref{fig:population} illustrates the exponential growth of the two modeled phyla, Firmicutes (Species 1) and Bacteroidetes (Species 2), alongside their combined total density $\rho(t)$. Both species display the characteristic lag, exponential, and early stationary phases of nutrient-limited batch culture. By the end of the simulation, Firmicutes attain roughly twice the density of Bacteroidetes, consistent with a higher intrinsic division rate. 

The corresponding autoinducer concentration $a(t)$ (Fig.~\ref{fig:autoinducer}) follows a sigmoidal trajectory, remaining near zero until the total cell density surpasses a critical threshold. 
This threshold behavior highlights how stochastic fluctuations in autoinducer production and diffusion can delay collective signaling until biomass accumulation reaches a sufficient level.

\begin{figure}[htp!]
\centering
\includegraphics[width=0.7\linewidth]{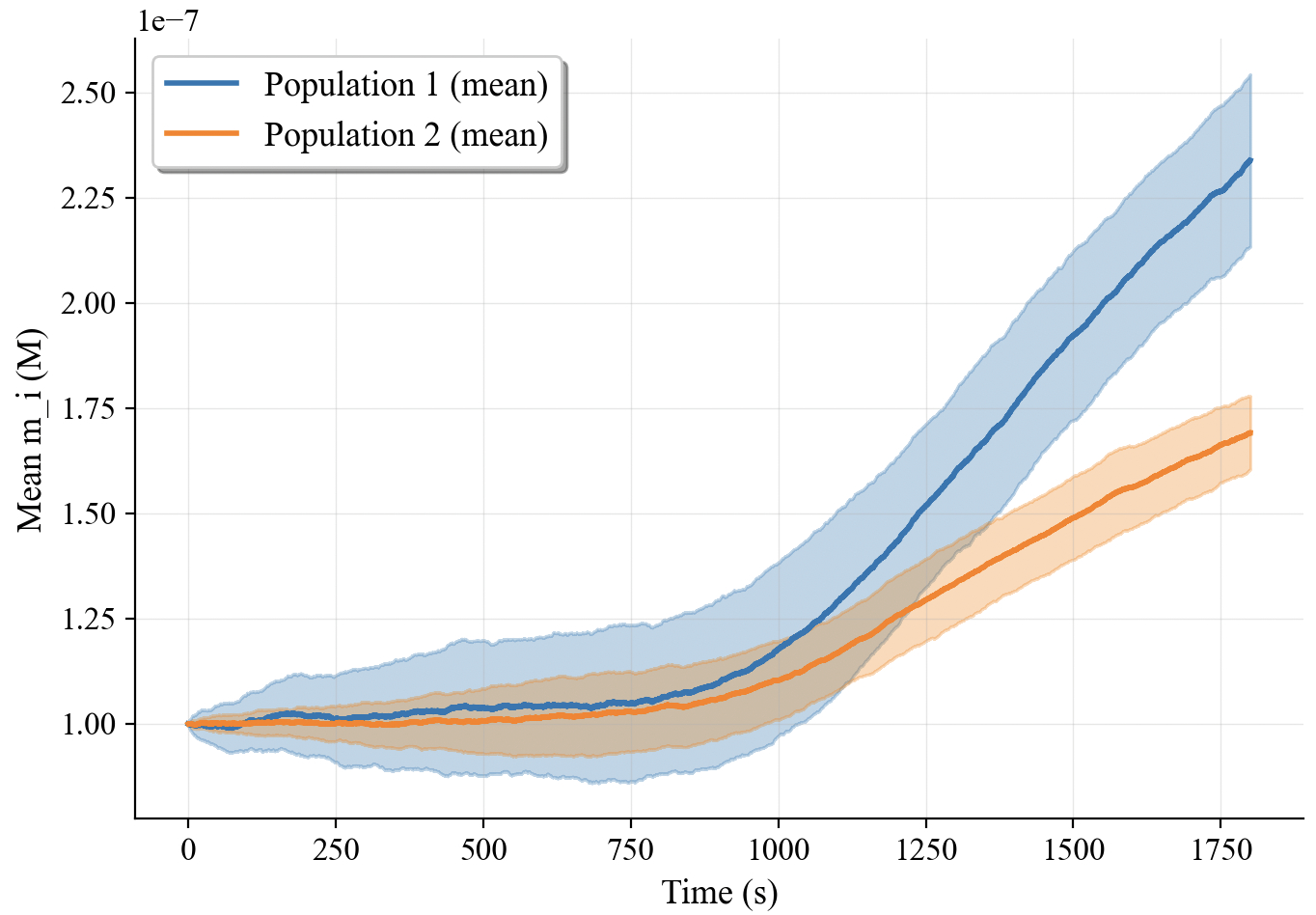}
\caption{Mean intracellular monitor protein concentration.}
\vspace{-1em}
\label{fig:mean_mi}
\end{figure}

\begin{figure}[htp!]
\centering
\includegraphics[width=1\linewidth]{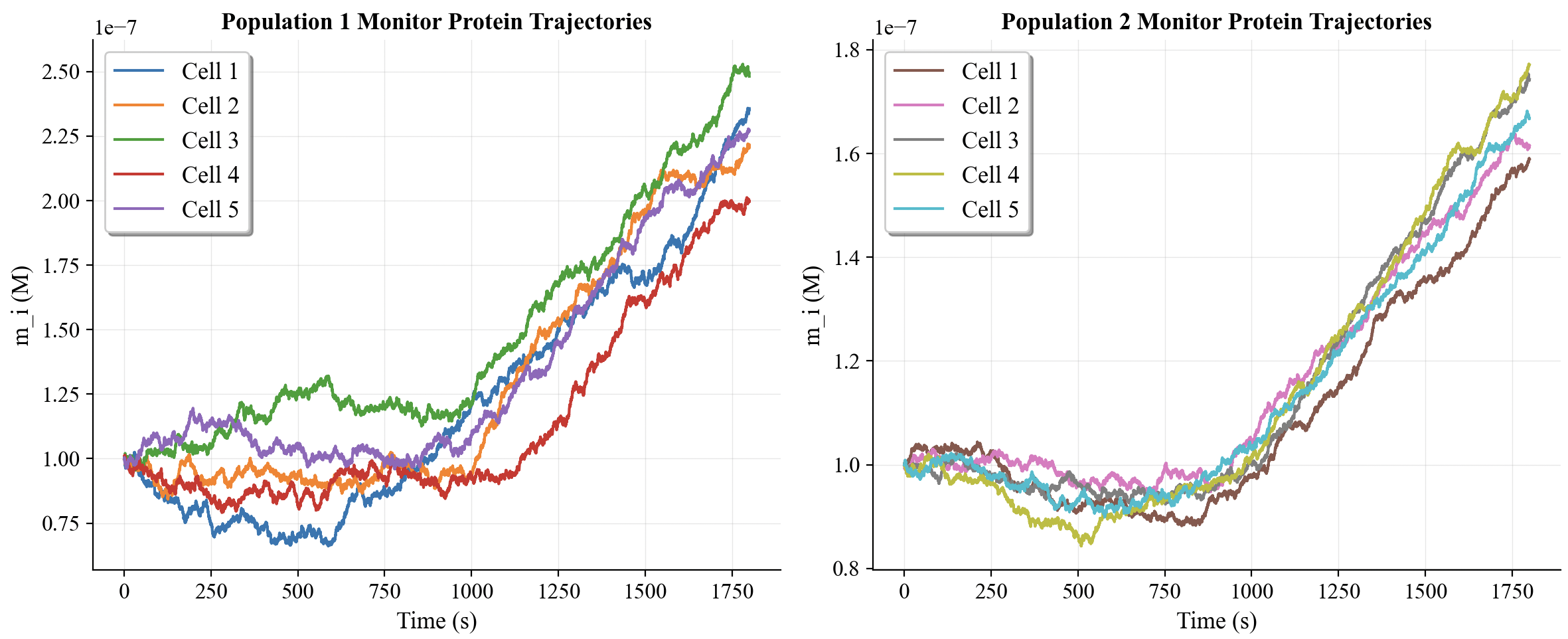}
\caption{Single-cell trajectories of monitor protein expression.}
\vspace{-1em}
\label{fig:trajectories}
\end{figure}

\paragraph{Monitor Protein Dynamics}
Fig.~\ref{fig:mean_mi} presents the mean intracellular concentration of the MP $m_i$ for each species, averaged over five replicates. Initially, both populations maintain a baseline expression.
Firmicutes show a steeper increase compared to Bacteroidetes. The error bands indicate greater variance among Firmicutes, suggesting that faster growth amplifies intrinsic expression noise. Single-cell trajectories of cells per species are given in Fig.~\ref{fig:trajectories}. These dynamics illustrate how bursty transcriptional events governs the fidelity of signal transduction.

\paragraph{Mutual Information}
Tab.~\ref{tab:regulation_capacities} provides mutual information between extracellular autoinducer levels $a(t)$ and intracellular MP levels $m(t)$ using a $k$-nearest-neighbor estimator with block bootstrap resampling to obtain 95\% confidence intervals. Two scenarios were compared: mutual information with cell-averaged responses ($\bar{m}$) and mutual information across all cells individually. Averaging across cells substantially increases information transfer, consistent with population-level noise buffering. Population~1 achieved $\approx 8.69$ bits (95\% CI: 8.69–8.78), while Population~2 reached $\approx 8.51$ bits (95\% CI: 8.52–8.62). In contrast, when considering single-cell outputs, information significantly dropped to around $2$ bits. This gap highlights the role of collective encoding in mitigating extrinsic fluctuations.  

\begin{table}[htp!]
\vspace{-1em}
\centering
\begin{tabular}{|l|c|}
\hline
\textbf{Condition} & \textbf{Mutual Information (bits)} \\
\hline
$MI(a;\bar{m}_{\text{pop1 avg}})$ & $8.692$ \\
$MI(a;\bar{m}_{\text{pop2 avg}})$ & $8.505$ \\
$MI(a;m_{\text{pop1}})$ & $1.786$ \\
$MI(a;m_{\text{pop2}})$ & $2.642$ \\
\hline
\end{tabular}
\caption{Empirical mutual information estimates. Averaging cells drastically increases transmitted information.}
\vspace{-1em}
\label{tab:regulation_capacities}
\end{table}

\paragraph{Noise Decomposition}

We partition output variance into intrinsic (MP stochasticity) and extrinsic (shared autoinducer fluctuations) contributions. It reveals that extrinsic fluctuations dominate both populations, accounting for $\sim72\%$ of variability, with intrinsic noise contributing $\sim28\%$. This imbalance suggests that collective encoding through shared extracellular signals introduces strong correlated noise that individual cells cannot filter alone.

\paragraph{Cross-species Asymmetry}

Cross-species information sharing was quantified via mutual information between mean MP levels and the partner’s cell density, as shown in Tab.\ref{tab:cross_species}. In the baseline mixed community (70--30 split), flow was nearly balanced with slight asymmetry favoring Population~1. Transfer entropy (TE) revealed weak directed influences, with $TE(\text{P2}\to\text{P1}) \approx 0.009$ bits and negligible reverse flow.  When Firmicutes dominated (90--10), asymmetry shifted toward Population~1 $\to$ Population~2, whereas a Bacteroidetes-dominant case (40--60) reversed the direction, with $TE(\text{P2}\to\text{P1}) \approx 0.025$ bits. Our results suggest that altered F/B balance not only changes population size but also reshapes information flow.

\begin{table}[htp!]
\vspace{-1em}
\centering
\begin{tabular}{|l|c|c|}
\hline
\textbf{Condition} & $\mathbf{I[\bar{m}_{\text{pop1}};\rho_{\text{pop2}}]}$ \textbf{(bits)} & $\mathbf{I[\bar{m}_{\text{pop2}};\rho_{\text{pop1}}]}$ \textbf{(bits)} \\
\hline
Baseline (current sim) & $8.64$  & $8.44 $ \\
Firmicutes-dominant &  8.71 & 7.80 \\
Bacteroidetes-dominant & 8.46 & 8.69 \\
\hline
\end{tabular}
\caption{Cross-species information transfer under baseline and asymmetric parameter regimes. Current simulations indicate slight asymmetry favoring Pop2 $\to$ Pop1 directionality.}
\vspace{-1em}
\label{tab:cross_species}
\end{table}

\paragraph{Temporal Information Processing}

\begin{figure}[htp!]
\vspace{-1em}
\centering
\includegraphics[width=1\linewidth]{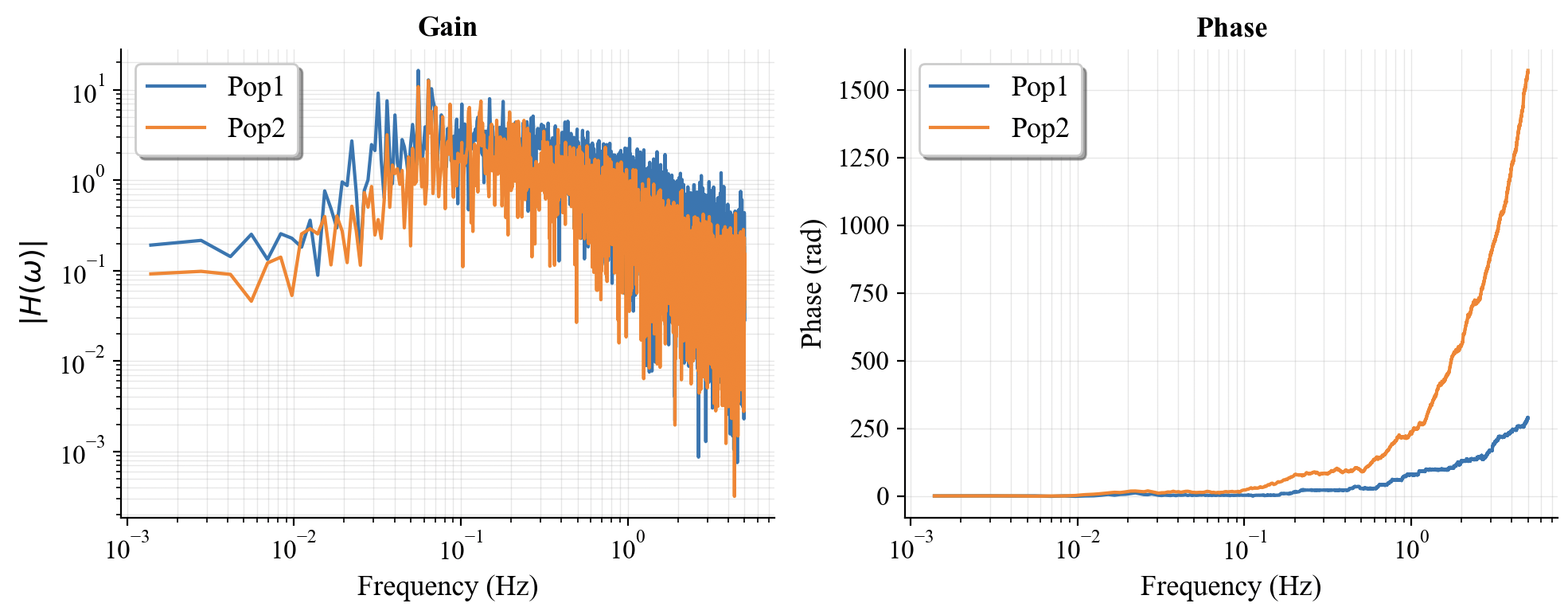}
\caption{Temporal frequency response and information transmission efficiency.}
\vspace{-1em}
\label{fig:frequency}
\end{figure}

Dynamic information flow was analyzed using empirical transfer function estimates from sinusoidal perturbations of the shared autoinducer. Bode plots (Fig.~\ref{fig:frequency}) show that both populations act as low-pass filters, transmitting low-frequency fluctuations efficiently while attenuating high-frequency inputs. The gain falls off sharply beyond $\sim 0.1$ Hz, and phase lags accumulate substantially at higher frequencies, with Population~2 showing larger delays. These results highlight strong temporal constraints in the signaling architecture.

\paragraph{Sensitivity Analysis}

We computed elasticities of average mutual information $I(a;\bar m)$ by perturbing LuxS secretion, autoinducer turnover $\tau_a$, and monitor degradation $\tau_\delta$ by $\pm 20\%$ and re–simulating. The sensitivity index is 
$S = \tfrac{I_{+} - I_{-}}{2 \epsilon I_{0}}, \ \epsilon = 0.2$, 
with $I_{0}$ the baseline.

Indices were small: LuxS secretion $S \approx -0.002$, $\tau_a \approx -0.015$, and $\tau_\delta \approx +0.001$. Thus, encoding operates near saturation: faster $\tau_a$ weakly lowers information transfer, while LuxS and $\tau_\delta$ effects were negligible. Overall, the channel is buffered against moderate parameter changes, with capacity shaped more by architecture than rate tuning.

\section{Conclusions}

We showed that gut microbial QS is shaped primarily by autoinducer noise, buffered through population averaging but limiting single-cell fidelity. Information flow is asymmetric, with the dominant group exerting stronger influence and the system is robust to moderate parameter changes. Given the importance of Firmicutes and Bacteroidetes in intestinal homeostasis, impaired information processing may underlie dysregulated microbial communication in disorders. These findings provide a quantitative basis to understand microbial communication and suggest extensions to multi-signal interactions, spatial effects, and host-driven extrinsic fluctuations.


\bibliographystyle{IEEEtran}
\bibliography{references.bib}

\end{document}